\documentstyle[epsf]{article}

\begin{document}

\renewcommand\baselinestretch{0.9}
\large\normalsize

\title{Jordan-Wigner fermions
       \protect\\
       and the spin-$\frac{1}{2}$ anisotropic $XY$ model 
       on a square lattice}

\author{Oleg Derzhko$^{a,b}$,
Johannes Richter$^c$
and Taras Verkholyak$^a$\\
{\small {$^a$Institute for Condensed Matter Physics,}}\\
{\small {1 Svientsitskii Street, L'viv-11, 79011, Ukraine}}\\
{\small {$^b$Chair of Theoretical Physics,
Ivan Franko National University of L'viv,}}\\
{\small {12 Drahomanov Street, L'viv-5, 79005, Ukraine}}\\
{\small {$^c$Institut f\"ur Theoretische Physik,
Universit\"at Magdeburg,}}\\
{\small {P.O. Box 4120, D-39016 Magdeburg, Germany}}}

\date{\today}

\maketitle

\begin{abstract}
Using the two-dimensional Jordan-Wigner fermionization
we calculate the thermodynamic quantities 
of the (spatially anisotropic) square-lattice 
spin-$\frac{1}{2}$ anisotropic $XY$ ($XZ$) model.
We compare the results of different approaches 
for the ground-state and thermodynamic properties of the model. 
\end{abstract}

\vspace{5mm}

\noindent
{\bf {PACS number(s):}}
75.10.-b

\vspace{5mm}


\vspace{5mm}

\noindent
{\bf {Postal address:}}\\
Dr. Oleg Derzhko (corresponding author)\\
Institute for Condensed Matter Physics\\
1 Svientsitskii Street, L'viv-11, 79011, Ukraine\\
tel/fax: (0322) 76 19 78\\
email: derzhko@icmp.lviv.ua

\vspace{5mm}

\renewcommand\baselinestretch{1.1}
\large\normalsize

Two-dimensional (2D) quantum spin models 
have been extensively studied during last years 
mainly because it is believed 
that they may be of use 
for describing the magnetic properties of CuO$_2$ layers 
in the high-temperature superconductors \cite{001}.
There exist a number of analytical approaches 
for a study of the thermodynamic properties 
of 2D quantum spin models,
e.g.,
the conventional spin-wave analysis,
the Green function technique, 
the approach based on the 2D Jordan-Wigner fermionization
as well as the coupled cluster method, 
the correlated basis function method 
etc..
In what follows 
we consider 
the spin-$\frac{1}{2}$ anisotropic $XY$ model 
on a spatially anisotropic square lattice 
within the framework of the scheme 
based on the 2D Jordan-Wigner fermionization 
and compare the results derived 
for the ground-state and thermodynamic quantities
with the exact ones
(1D limit, square-lattice Ising model) 
and the predictions of other approximate theories.
The performed calculations 
yield an impression  
about the region of validity of some approaches 
usually applied for a study of thermodynamics 
of 2D quantum spin models.

We start from a model of $N\to\infty$ spins $\frac{1}{2}$ 
on a spatially anisotropic square lattice
governed by the anisotropic $XY$ Hamiltonian
\begin{eqnarray}
\label{001}
H=\sum_{i=0}^{\infty}\sum_{j=0}^{\infty}
\biggl(
J\Bigl((1+\gamma)s^x_{i,j} s^x_{i+1,j}
+(1-\gamma)s^y_{i,j} s^y_{i+1,j}\Bigr)
\biggr.
\nonumber\\
\biggl.
+J_{\perp}\Bigl((1+\gamma)s^x_{i,j} s^x_{i,j+1}
+(1-\gamma)s^y_{i,j} s^y_{i,j+1}\Bigr)
\biggr).
\end{eqnarray}
Here $J$ and $J_{\perp}=RJ$ 
are the exchange interactions
between the neighbouring sites
in a row and a column, respectively
(for concreteness we assume both to be positive),
and the parameter $\gamma$ 
controls the anisotropy of the exchange interaction.
Making use of the 2D Jordan-Wigner fermionization 
and adopting a mean-field treatment of the phase factors
which appear after the fermionization \cite{002,003}
we perform consequently 
the Fourier and Bogolyubov transformations 
to arrive at the following Hamiltonian of noninteracting spinless fermions 
which represent the initial spin model (\ref{001})
\begin{eqnarray}
\label{002}
H=\sum_{{\bf{k}}}\nolimits^{\prime}
\sum_{\alpha=1}^2
\Lambda_{\alpha}({\bf{k}})
\left(\eta^+_{{\bf{k}},\alpha}\eta_{{\bf{k}},\alpha}
-\frac{1}{2}\right),
\\
\Lambda_1({\bf{k}})
=\sqrt{\left(J_{\perp}\cos k_y+\gamma J\cos k_x\right)^2
+\left(J\sin k_x+\gamma J_{\perp}\sin k_y\right)^2},
\nonumber\\
\Lambda_2({\bf{k}})
=\sqrt{\left(J_{\perp}\cos k_y-\gamma J\cos k_x\right)^2
+\left(J\sin k_x-\gamma J_{\perp}\sin k_y\right)^2}
\nonumber
\end{eqnarray}
(the prime denotes 
that ${\bf{k}}$ in the thermodynamic limit varies in the region 
$-\pi\le k_x\le\pi$,
$-\pi+\vert k_x\vert\le k_y\le\pi-\vert k_x\vert$).
The Helmholtz free energy per site
\begin{eqnarray}
\label{003}
f=-\frac{1}{2\beta}
\int_{-\pi}^{\pi}\frac{{\mbox{d}}k_x}{2\pi}
\int_{-\pi}^{\pi}\frac{{\mbox{d}}k_y}{2\pi}
\left(
\ln\left(2\cosh\frac{\beta\Lambda_1({\bf{k}})}{2}\right)
+
\ln\left(2\cosh\frac{\beta\Lambda_2({\bf{k}})}{2}\right)
\right)
\end{eqnarray}
yields the thermodynamic properties of the spin model (\ref{001}).
In Fig. 1 we plot the ground-state energy per site
of the spin model (\ref{001}), (\ref{002}) (dotted curves) 
in comparison with the exact results 
if $R=0$ (1D $XY$ model)
or $\gamma=1$ (square-lattice Ising model)
and the spin-wave theory result for $\gamma=0$, $R=1$
(spatially isotropic square-lattice isotropic $XY$ model).
Eq. (\ref{003}) contains the exact result in 1D limit (Fig. 1b),
however, 
deviates noticeably from the exact result for $\gamma=1$
(compare the curves 3 in Fig. 1a).
For $\gamma=0$, $R=1$ 
Eq. (\ref{003}) yields the result 
which differs from the spin-wave theory prediction 
denoted by the full circles.
(The outcomes of different numerical approaches 
(see \cite{004})
lie within the full circles.) 
From the exact calculation for $\gamma=1$ \cite{005} 
we know that the temperature dependence of the specific heat 
exhibits a logarithmic singularity.
Obviously,
the Jordan-Wigner fermions (\ref{002}), (\ref{003})
cannot reproduce this peculiarity 
inherent in the spin model.

It is worth to remind here
that the conventional spin-wave theory 
was originally thought to be unsatisfactory 
for quantum $XY$ models \cite{006}.
However,
the authors of the paper \cite{007} showed
that considering the $XZ$ rather than the $XY$ Hamiltonian 
one gets within the spin-wave theory satisfactory results 
of the same quality  
as for the Heisenberg Hamiltonian.
Following this idea 
we perform the rotation of the spin axes 
$s^x\to -s^z$,
$s^y\to s^x$,
$s^z\to -s^y$
and consider instead of (\ref{001}) 
the following Hamiltonian 
\begin{eqnarray}
\label{004}
H=\sum_{i=0}^{\infty}\sum_{j=0}^{\infty}
\biggl(
J\Bigl((1-\gamma)s^x_{i,j} s^x_{i+1,j}
+(1+\gamma)s^z_{i,j} s^z_{i+1,j}\Bigr)
\biggr.
\nonumber\\
\biggl.
+
J_{\perp}\Bigl((1-\gamma)s^x_{i,j} s^x_{i,j+1}
+(1+\gamma)s^z_{i,j} s^z_{i,j+1}\Bigr)
\biggr).
\end{eqnarray}
Proceeding further with (\ref{004}) in the described above manner 
and assuming (for concreteness) 
antiferromagnetic long-range order
while decoupling the quartic fermionic terms \cite{008}
we get instead of (\ref{002}) 
the following Hamiltonian 
\begin{eqnarray}
\label{005}
H=\sum_{{\bf{k}}}\nolimits^{\prime}
\sum_{\alpha=1}^2
\Lambda_{\alpha}({\bf{k}})
\left(\eta^+_{{\bf{k}},\alpha}\eta_{{\bf{k}},\alpha}
-\frac{1}{2}\right)
+N\left(1+\gamma\right)\left(J+J_{\perp}\right)m^2,
\\
\Lambda_{1,2}({\bf{k}})
=2\sqrt{{\cal{A}}^2+{\cal{B}}^2+{\cal{C}}^2+{\cal{D}}^2+{\cal{M}}^2
\pm 2\sqrt{2{\cal{A}}{\cal{B}}{\cal{C}}{\cal{D}}
+{\cal{A}}^2{\cal{D}}^2
+{\cal{B}}^2{\cal{C}}^2
+{\cal{M}}^2\left({\cal{B}}^2+{\cal{D}}^2\right)
}},
\nonumber\\
{\cal{A}}=\frac{1-\gamma}{4}J_{\perp}\cos k_y,\;\;\;
{\cal{B}}=\frac{1-\gamma}{4}J_{\perp}\sin k_y,
\nonumber\\
{\cal{C}}=\frac{1-\gamma}{4}J\sin k_x,\;\;\;
{\cal{D}}=\frac{1-\gamma}{4}J\cos k_x,
\nonumber\\
{\cal{M}}=\left(1+\gamma\right)\left(J+J_{\perp}\right)m,
\nonumber
\end{eqnarray}
where $m$ is determined self-consistently 
by minimizing the Helmholtz free energy per site
\begin{eqnarray}
\label{006}
2\left(1+\gamma\right)\left(J+J_{\perp}\right)m
=\frac{1}{4}
\int_{-\pi}^{\pi}
\int_{-\pi}^{\pi}
\frac{{\mbox{d}}{\bf{k}}}{\left(2\pi\right)^2}
\left(
\frac{\partial\Lambda_1({\bf{k}})}{\partial m}
\tanh\frac{\beta\Lambda_1({\bf{k}})}{2}
+
\frac{\partial\Lambda_2({\bf{k}})}{\partial m}
\tanh\frac{\beta\Lambda_2({\bf{k}})}{2}
\right).
\end{eqnarray}
In Fig. 2 we plot the ground-state energy 
of the spin model (\ref{004}), (\ref{005}), (\ref{006}) 
(dashed curves). 
The results based on Eqs. (\ref{005}), (\ref{006}) for $\gamma=1$
reproduce the exact result 
for square-lattice Ising model 
(curve 3 in Fig. 2a)
as well as the spin-wave theory prediction 
for $\gamma=0$, $R=1$.
However, 
the result based on Eqs. (\ref{005}), (\ref{006}) 
for $R=0$
does not coincide with the exact one 
in 1D limit  
(curve 1 in Fig. 2b).

To summarize,
we have calculated the thermodynamic quantities 
for the spin-$\frac{1}{2}$ anisotropic $XY$ ($XZ$) model 
on a spatially anisotropic square lattice  
using the 2D Jordan-Wigner fermionization.
To reveal a quality of the results 
obtained within the framework of this approach 
we have compared them with the exact results 
available in 1D limit and extremely anisotropic exchange interaction limit
(Ising interaction).
We have found 
that although 
there is an agreement with the spin-wave theory 
and other approximate approaches 
a disagreement with the exact results may be noticeable. 
Thus, the question about the quality 
of the results based on the 2D Jordan-Wigner fermionization
(as well as, e.g.,  of the spin-wave theory results) 
remains still open
and requires further studies. 
Moreover,
for arbitrary values of anisotropy parameter 
$\gamma$ ($\gamma\ne 1$) for $R\ne 0$ 
a comparison with the exact diagonalization data 
and results of other numerical approaches 
is desirable.

\vspace{5mm}

The authors thank T. Krokhmalskii
for helpful discussions on this topic. 
The present study was partly supported by the DFG project 436 UKR 17/101.
O. D. acknowledges 
the kind hospitality of the Magdeburg University in the summer of 2001 
when this paper was completed.

\vspace{5mm}

FIGURE CAPTURES

\vspace{2mm}

FIGURE 1.
The ground-state energy per site 
for the square-lattice spin-$\frac{1}{2}$ anisotropic $XY$ model 
(\ref{001})
$e_0$ vs. $R$ (a)
(1 -- $\gamma=0$,
2 -- $\gamma=0.5$,  
3 -- $\gamma=1$) 
and
$e_0$ vs. $\gamma$ (b)
(1 -- $R=0$,
2 -- $R=0.5$,  
3 -- $R=1$);
exact results (solid curves) 
and
the approximate results 
obtained on the basis of (\ref{002})
(dotted curves);
the full circles correspond to the spin-wave result for $\gamma=0$, $R=1$.

\vspace{2mm}

FIGURE 2.
The same as in Fig. 1 
for the $XZ$ Hamiltonian (\ref{004});
the approximate results 
obtained on the basis of (\ref{005}), (\ref{006})
are shown by dashed curves.

\clearpage

\begin{figure}[t]
\epsfysize=150mm
\epsfclipon
\centerline{\epsffile{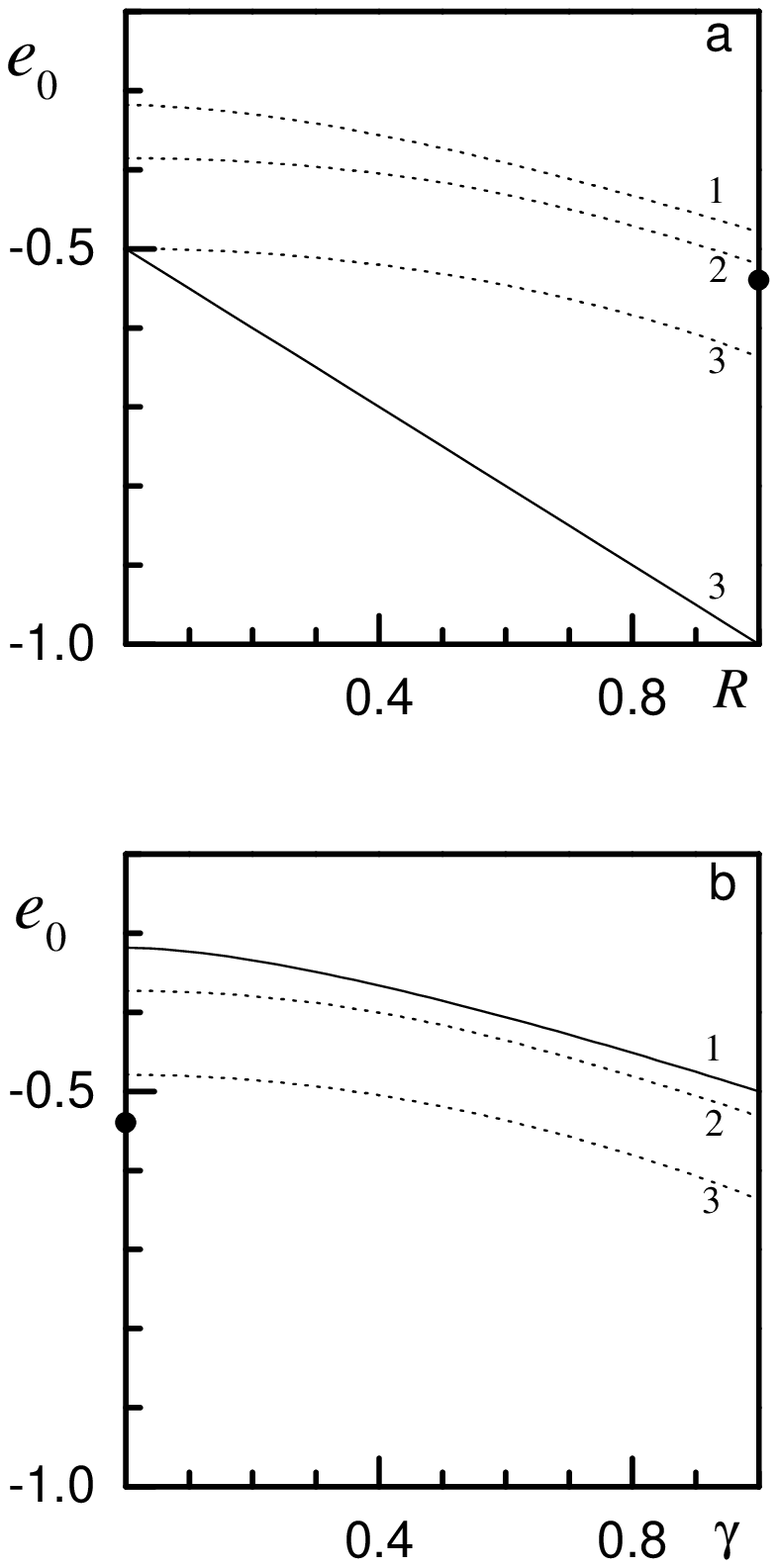}}
\vspace{30mm}
\caption[]
{\small }
\label{fig1}
\end{figure}

\clearpage

\begin{figure}[t]
\epsfysize=150mm
\epsfclipon
\centerline{\epsffile{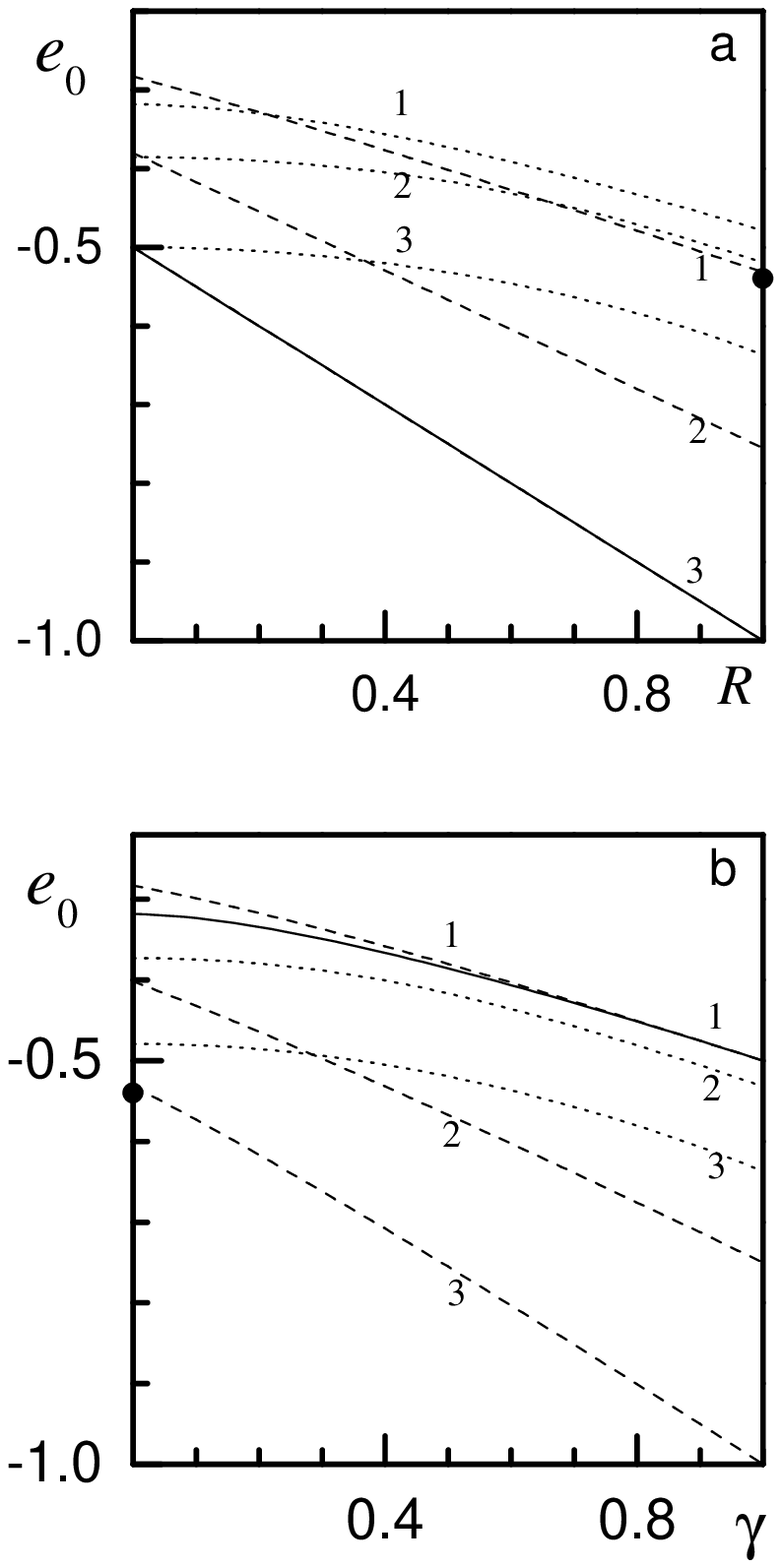}}
\vspace{30mm}
\caption[]
{\small }
\label{fig3}
\end{figure}


\begin{thebibliography}{}

\bibitem{001}
  M. A. Kastner, R. J. Birgeneau, G. Shirane,
  and Y. Endoh,
  Rev. Mod. Phys. {\bf 70,} 897 (1998).
 
\bibitem{002}
  Y. R. Wang,
  Phys. Rev. B {\bf 43,} 3786 (1991).

\bibitem{003}
  O. Derzhko,
  to appear in
  J. Phys. Stud. (L'viv) {\bf 5,} (2001)
  (cond-mat/0101188).

\bibitem{004}
  E. Loh, Jr., D. J. Scalapino, and P. M. Grant,
  Phys. Rev. B {\bf 31,} 4712 (1985);\\  
  C. J. Hamer, T. H\"{o}velborn, and M. Bachhuber,
  J. Phys. A {\bf 32,} 51 (1999);\\
  D. J. J. Farnell and M. L. Ristig,
  cond-mat/0105386.

\bibitem{005}
  D. C. Mattis,
  The Theory of Magnetism II.
  Thermodynamics and Statistical Mechanics
  (Springer-Verlag:
  Berlin Heidelberg New York Tokyo, 1985).  

\bibitem{006}
  D. C. Mattis,
  The Theory of Magnetism I.
  Statics and Dynamics
  (Springer-Verlag:
  Berlin Heidelberg New York, 1981), Chapter 5.

\bibitem{007} 
  G. Gomez-Santos and J. D. Joannopoulos,
  Phys. Rev. B {\bf 36,} 8707 (1987).
  
\bibitem{008}
  The existence of long-range order 
  was rigorously proved by
  T. Kennedy, E. H. Lieb, and B. S. Shastry,
  Phys. Rev. Lett. {\bf 61,} 2582 (1988).

\end{thebibliography}
\end{document}